\documentclass[12pt,journal]{IEEEtran}
\usepackage{amsmath}
\usepackage{graphicx}
\usepackage{mathtools}
\usepackage{amsfonts}
\usepackage{pifont}
\usepackage{amssymb}
\usepackage{epstopdf}
\usepackage{color}
\usepackage[utf8]{inputenc}
\usepackage{setspace}
\usepackage{ragged2e}
\usepackage{epsfig}
\usepackage{algorithmic}
\usepackage{textcomp}

\makeatletter
\newcommand{\vast}{\bBigg@{1.2}}
\newcommand{\Vast}{\bBigg@{2.5}}
\newcommand{\vastl}{\bBigg@{3.9}}
\newcommand{\Vastl}{\bBigg@{2.1}}
\makeatother

\onecolumn
\begin{document}

\title{Secrecy Analysis of Physical Layer over $\kappa-\mu$ Shadowed Fading Scenarios}
\author{\vspace{0.5 cm}\IEEEauthorblockN{Hussien Al-Hmood, \textit{Member, IEEE,} and H. S. Al-Raweshidy, \textit{Senior Member, IEEE}}}
\maketitle

\begin{abstract}
In this paper, the secrecy analysis of physical layer when both the main and wiretap channels undergo $\kappa-\mu$ shadowed fading channel is investigated. In particular, the average secrecy capacity (ASC), secure outage probability (SOP), the lower bound of SOP (SOP$^L$), and the probability of strictly positive secrecy capacity (SPSC) are derived by using the classic Wyner’s wiretap model. Two different scenarios for the fading parameters, i.e., $\mu$ and $m$ which represents the shadowing impact have been studied. These parameters are chosen first as arbitrary numbers, thus the performance metrics are expressed in single infinite series with multivariate Meijer $G$-function. In the second scenario, both the aforementioned fading parameters are assumed to be integer numbers in order to obtain the derived results in simple exact closed-form analytic mathematically tractable expressions. The numerical results of this analysis are
 verified via Monte Carlo simulations.
\end{abstract}


\section{Introduction}
Wyner has developed the information-theoretic notion of perfect secrecy that was introduced by Shannon via proposing the wiretap channel. The notion of this channel includes a legitimate user communicates with the intended receiver which are named Alice and Bob, respectively, in the presence of an eavesdropper [1]. Accordingly, the performance analysis of the physical layer security over fading channels has been given a special attention in the recent works. For example, the probability of strictly positive secrecy capacity (SPSC), the secure outage probability (SOP), and the average secrecy capacity (ASC) when the wireless channels undergo the additive white Gaussian noise (AWGN) and the Rayleigh fading channel are derived in [2] and [3]. The SPSC when both the main and wiretap channels undergo Rician fading scenario is given in [4]. The SOP and SPSC over Rician/Nakagami-m and Nakagami-m/Rician fading scenarios are provided in [5]. The SPSC and the ASC of the Weibull fading channel are introduced in [6] and [7], respectively. The closed-form expression of SPSC when both the Bob and the eavesdropper experience log-normal fading is presented in [8].
\par Recently, many efforts have been devoted to study the secrecy performance of the physical layer over different generalized fading channels which give results closer to the practical measurement than the traditional distributions. For instance, the closed-form expressions for the SPSC and the lower bound of SOP (SOP$^L$) over generalised Gamma fading model are derived by using the classic Wyner’s wiretap model [9]. The analysis in [10] is investigated by analysing the SPSC and SOP$^L$ over $\kappa-\mu$ fading that is used to model the line-of-sight (LoS) communication scenario where the parameters $\kappa$ and $\mu$ denote the ratio between the powers of the dominant and the scattered waves components and the number of multipath clusters, respectively. In [11] and [12], the SPSC and the ASC are, respectively, utilised to analyse the secrecy performance over $\alpha-\mu$ fading which is proposed to represent the non-homogeneous environment of wireless channel where $\alpha$ indicates the non-linearity index. The scenarios of mean/wiretap channels undergo $\alpha-\mu$/$\kappa-\mu$ and $\kappa-\mu/\alpha-\mu$ fading conditions are given in [13] to study the secrecy capacity of physical layer via deriving the expression of the SPSC, the SOP, and the SOP$^L$.   
\par The wireless channel may subject to the shadowing effect which is part of fading that can not be ignored. Hence, several works have been dedicated to analyse the security of physical layer over composite multipath/shadowed fading channels. The generalised-$K$ ($GK$) fading model which is composite of Nakagami-$m$/log-normal is employed to represent the main/wiretap channels of the classic Wyner's framework in [14] and [15]. In the former, the ASC, the SOP, and the SPSC are expressed in terms of the extended generalized bivariate Meijer $G$-function (EGBMGF) whereas a mixture Gamma distribution is used in the latter to approximate with high accuracy the same performance metrics. The analysis in [16] is based on using the $\kappa-\mu$ shadowed fading channels to derive the SPSC, and the SOP$^L$ of physical layer. However, the provided expressions are approximated and included the EGBMGF as well as two infinite series that are not easily and steadily convergence. Therefore, the authors have used the Gamma distribution to approximate the expressions of the aforementioned performance metrics in simple closed-from formats.
\par Unlike [16], this work analyses the secrecy performance of the physical layer over $\kappa-\mu$ shadowed fading channel via different formats of the ASC, the SOP, the SOP$^L$, and the SPSC. The aforementioned metrics are expressed first in single infinite series and the EGBMGF which would make the convergence acceleration is faster in comparison with [16]. In the second scenario, the parameter $\mu$ and the shadowing severity index are supposed to be integer numbers. Consequently, the derived results are given in simple exact analytic mathematically tractable closed-from expressions. Furthermore, in [16], an approximation is utilised to obtain the closed-from expression of the SOP. It is remarkable that the ASC and the SOP in $\kappa-\mu$ shadowed fading have not been provided in [16].
\par \textit{Organization:} Section II describes the general and limited formats of the probability density function (PDF) and the cumulative distribution function of $\kappa-\mu$ shadowed fading. The ASC, the SOP, the SOP$^L$, and the SPSC for general and integer values of $\mu$ and shadowing parameters are derived in Sections III, IV, V, and VI, respectively. Section VII performs the Monte Carlo simulation and numerical results. Finally, conclusions are presented in Section VIII. 
\section{The PDF and CDF of $\kappa-\mu$ Shadowed Fading}
The PDF of the instantaneous SNR $\gamma_i$, $f_{\gamma_i}(\gamma_i)$, for the destination (Bob), $D$, and the eavesdropper, $E$, channels in $\kappa-\mu$ shadowed fading model is given by [17, eq. (4)]
\setcounter{equation}{0}
\label{eqn_1}
\begin{equation}
f_{\gamma_i}(\gamma_i)=\Theta_i \gamma^{\mu_i-1}_i e^{-\mathcal{A}_i \gamma_i} {_1F_1}(m_i;\mu_i;\mathcal{B}_i \gamma_i).
\end{equation}
where $i \in \{D, E\}$, $\Theta_i=\frac{\mu^{\mu_i}_i m^{m_i}_i (1+\kappa_i)^{\mu_i}}{\Gamma(\mu_i) \bar{\gamma}^{\mu_i}_i (\mu_i \kappa_i+m_i)^{m_i}}$, $\mathcal{A}_i = \frac{\mu_i(1+\kappa_i)}{\bar{\gamma}_i}$, $\mathcal{B}_i=\frac{\mu^2_i \kappa_i (1+\kappa_i)}{(\mu_i \kappa_i+m_i) \bar{\gamma}_i}$, $\bar{\gamma}_i$ is the average SNR, $m_i$ is the shadowing severity index, and ${_1F_1}(.;.;.)$ is the confluent hypergeometric function defined in [18, eq. (9.14.1)].  
\par The CDF of the $\kappa-\mu$ shadowed fading channel is expressed as [17, eq. (6)] 
\setcounter{equation}{1}
\label{eqn_2}
\begin{equation}
F_{\gamma_i}(\gamma_i)=\frac{\Theta_i}{\mu_i}\gamma^{\mu_i}_i\Phi_2\big(\mu_i-m_i,m_i;\mu_i-1;-\mathcal{A}_i \gamma_i,\mathcal{C}_i \gamma_i \big).
\end{equation}
where $\mathcal{C}_i=\frac{m_i}{\mu_i \kappa_i+m_i}\mathcal{A}_i$, $\Phi_2(.,.;.;.,.)$ is the bivariate confluent hypergeometric function [18, eq. (9.261.2)].
\par When both $\mu$ and $m$ are integer numbers, i.e., $\mu$ and $m$ $\in$ $\mathbb{Z}^+$, the PDF and the CDF are, respectively, given by [19, eqs. (12) and eq. (13)]
\label{eqn_3}
\begin{equation}
f_{\gamma_i}(\gamma_i)=\sum_{j_i=0}^{M_i}\Lambda_{j_i}\frac{\gamma^{\psi_{j_i}-1}_i}{\Omega^{\psi_{j_i}}_{j_i} (\psi_{j_i}-1)!}e^{-\frac{\gamma_i}{\Omega_{j_i}}}.
\end{equation}
and
\label{eqn_4}
\begin{equation}
F_{\gamma_i}(\gamma_i)= 1-\sum_{j_i=0}^{M_i}\Lambda_{j_i}e^{-\frac{\gamma_i}{\Omega_{j_i}}}\sum_{r_i=0}^{\psi_{j_i}-1}\frac{1}{r_i!}\bigg(\frac{\gamma_i}{\Omega_{j_i}}\bigg)^{r_i}.
\end{equation}
where $M_i$, $\Lambda_{j_i}$, $\psi_{j_i}$, and $\Omega_{j_i}$ are provided in Table I.
\label{Table_1}
\begin{table}[!t]
\renewcommand{\arraystretch}{1.1}
\caption{Parameter Values for the PDF and the CDF of the $\kappa-\mu$ Shadowed Fading with Integer $\mu$ AND $m$ [19].}
\centering
\begin{tabular}{c|c}
\hline
Case  &  Parameters \\
\hline 
\hline    & $M_i=\mu_i$ \\ 
$\mu_i > m_i$
&$\Lambda_{j_i}= \vastl\{\begin{array}{lr}
        0, \hspace{4.0 in} j_i=0\\
        (-1)^{m_i}{m_i+j_i-2 \choose j_i-1}\vast[\frac{\mathcal{C}_i}{\mathcal{A}_i}\vast]^{m_i}\vast[\frac{\mathcal{B}_i}{\mathcal{A}_i}\vast]^{-m_i-j_i+1}, \hspace{1.2 in} 0< j_i \leqslant \mu_i-m_i\\
        (-1)^{j_i-\mu_i+m_i-1}{j_i-2 \choose j_i-\mu_i+m_i-1}\vast[\frac{\mathcal{C}_i}{\mathcal{A}_i}\vast]^{j_i-\mu_i+m_i-1}\vast[\frac{\mathcal{B}_i}{\mathcal{A}_i}\vast]^{-j_i+1}, \hspace{0.3 cm} \mu_i-m_i< j_i \leqslant \mu_i
        \end{array} $    \\
                                  & $\psi_{j_i}= \Vastl\{\begin{array}{lr}
        \mu_i-m_i-j_i+1, & 0 \leqslant j_i \leqslant \mu_i-m_i\\
        \mu_i-j_i+1, & \mu_i-m_i< j_i \leqslant \mu_i
        \end{array} $ \\
                                 & $\Omega_{j_i}=\Vastl\{\begin{array}{lr}
        \frac{1}{\mathcal{A}_i}, & 0 \leqslant j_i \leqslant \mu_i-m_i\\
        \frac{1}{\mathcal{C}_i}, & \mu_i-m_i< j_i \leqslant \mu_i \vspace{0.1 cm}
        \end{array}$ \\ 
\hline  
  & $M_i=m_i-\mu_i$ \\
  $\mu_i\leqslant m_i$ & $\Lambda_{j_i}={m_i-\mu_i \choose j_i}\vast[\frac{\mathcal{C}_i}{\mathcal{A}_i}\vast]^{j_i}\vast[\frac{\mathcal{B}_i}{\mathcal{A}_i}\vast]^{m_i-\mu_i-j_i}$ \\
   & $\psi_{j_i}=m_i-j_i$ \\
  & $\Omega_{j_i}= \frac{1}{\mathcal{C}_{j_i}} $ \\
\hline
\end{tabular}
\end{table} 
\section{Average Secrecy Capacity}
The ASC can be calculated by $\bar{C}_s=I_1+I_2-I_3$ [15, eq. (6)] where $I_1$, $I_2$, and $I_3$ are expressed as
\label{eqn_5}
\begin{equation}
I_1=\int_0^\infty \text{ln}(1+\gamma_D)f_D(\gamma_D)F_E(\gamma_D)d\gamma_D.
\end{equation} 

\label{eqn_6}
\begin{equation}
I_2=\int_0^\infty \text{ln}(1+\gamma_E)f_E(\gamma_E)F_D(\gamma_E)d\gamma_E.
\end{equation}

\label{eqn_7}
\begin{equation}
I_3=\int_0^\infty \text{ln}(1+\gamma_E)f_E(\gamma_E)d\gamma_E.
\end{equation}
\par For arbitrary values of $\mu$ and $m$, $I_1$, $I_2$, and $I_3$ are, respectively, given by (8), (9), and (10)
\label{eqn_8}
\begin{align}
I_1=&\Theta_D \Theta_E\frac{\Gamma(\mu_D)\Gamma(\mu_E)}{\Gamma(\mu_D-m_D)\Gamma(\mu_E-m_E)\Gamma(m_E)}\sum_{j=0}^\infty\frac{\mathcal{C}^j_E}{j!}\nonumber\\
&\times G^{0,1:1,2:1,1:1,2}_{1,0:2,2:1,2:2,2} 
\Bigg( \begin{matrix}
  1-j-\mu_D-\mu_E\\
  -\\
\end{matrix} \Vast\vert
\begin{matrix}
  1,1\\
  1,0\\
\end{matrix} \Vast\vert
\begin{matrix}
  1-\mu_D+m_D\\
  0,1-\mu_D\\
\end{matrix} \Vast\vert
\begin{matrix}
  1-\mu_E+m_E,1-m_E\\
  0,-\mu_E-j\\
\end{matrix} \Vast\vert  \hspace{2 cm}\nonumber\\
&  \hspace{9 cm} \frac{1}{\mathcal{A}_D-\mathcal{B}_D},\frac{\mathcal{B}_D}{\mathcal{A}_D-\mathcal{B}_D},\frac{\mathcal{A}_E}{\mathcal{A}_D-\mathcal{B}_D}
\Bigg).
\end{align}

\label{eqn_9}
\begin{align}
I_2=&\Theta_E \Theta_D\frac{\Gamma(\mu_E)\Gamma(\mu_D)}{\Gamma(\mu_E-m_E)\Gamma(\mu_D-m_D)\Gamma(m_D)}\sum_{j=0}^\infty\frac{\mathcal{C}^j_D}{j!}\nonumber\\
&\times G^{0,1:1,2:1,1:1,2}_{1,0:2,2:1,2:2,2} 
\Bigg( \begin{matrix}
  1-j-\mu_E-\mu_D\\
  -\\
\end{matrix} \Vast\vert
\begin{matrix}
  1,1\\
  1,0\\
\end{matrix} \Vast\vert
\begin{matrix}
  1-\mu_E+m_E\\
  0,1-\mu_E\\
\end{matrix} \Vast\vert
\begin{matrix}
  1-\mu_D+m_D,1-m_D\\
  0,-\mu_D-j\\
\end{matrix} \Vast\vert  \hspace{2 cm}\nonumber\\
&  \hspace{9 cm} \frac{1}{\mathcal{A}_E-\mathcal{B}_E},\frac{\mathcal{B}_E}{\mathcal{A}_E-\mathcal{B}_E},\frac{\mathcal{A}_D}{\mathcal{A}_E-\mathcal{B}_E}
\Bigg).
\end{align}

\label{eqn_10}
\begin{align}
I_3=\Theta_E \frac{\Gamma(\mu_E)}{\Gamma(\mu_E-m_E)}G^{0,1:1,2:1,1}_{1,0:2,2:1,2} 
\Bigg( \begin{matrix}
  1-\mu_E\\
  -\\
\end{matrix} \Vast\vert
\begin{matrix}
  1,1\\
  1,0\\
\end{matrix} \Vast\vert
\begin{matrix}
  1-\mu_E+m_E\\
  0,1-\mu_E\\
\end{matrix}\Vast\vert\frac{1}{\mathcal{A}_E-\mathcal{B}_E},\frac{\mathcal{B}_E}{\mathcal{A}_E-\mathcal{B}_E}
\Bigg). 
\end{align}

where $\Gamma(a) = \int_0^\infty x^{a-1} e^{-x} dx$ and $G(.)$ are the incomplete Gamma function and the EGBMGF as in [20, Table I], respectively. It can be noticed that the EGBMGF is not yet implemented in MATLAB and MATHEMATICA software packages. Therefore, a MATHEMATICA code that is available in [20] has been used in this work.  

\label{App_A}
\begin{IEEEproof}
See Appendix A.
\end{IEEEproof}

\par When both $\mu$ and $m$ are integer numbers, (5), (6), and (7) can be yielded in simple exact closed-form as follows
\label{eqn_11}
\begin{align}
I_1&=\sum_{j_D=0}^{M_D}\Lambda_{j_D}\bigg[e^{\mathcal{C}_{j_D}}\sum_{k=1}^{\psi_{j_D}}\frac{\Gamma(k-\psi_{j_D},\mathcal{C}_{j_D})}{\Omega^{\psi_{j_D}-k}_{j_D}}\nonumber\\
&-\sum_{j_E=0}^{M_E}\Lambda_{j_E}e^{\mathcal{C}_{j_D}+\mathcal{C}_{j_E}}\sum_{r_E=0}^{\psi_{j_E}-1}\frac{(\psi_{j_D})_{r_E}}{r_E!} \sum_{l=1}^{\psi_{j_D}+r_E} \frac{\Gamma(l-\psi_{j_D}-r_E, \mathcal{C}_{j_D}+\mathcal{C}_{j_E})}{\Omega^{\psi_{j_D}-l}_{j_D}\Omega_{j_E}^{r_E-l}(\Omega^{\psi_{j_D}}+\Omega^{\psi_{j_E}})^l}\bigg].
\end{align} 

\label{eqn_12}
\begin{align}
I_2&=\sum_{j_E=0}^{M_E}\Lambda_{j_E}\bigg[e^{\mathcal{C}_{j_E}}\sum_{k=1}^{\psi_{j_E}}\frac{\Gamma(k-\psi_{j_E},\mathcal{C}_{j_E})}{\Omega^{\psi_{j_E}-k}_{j_E}}\nonumber\\
&-\sum_{j_D=0}^{M_D}\Lambda_{j_D}e^{\mathcal{C}_{j_E}+\mathcal{C}_{j_D}}\sum_{r_D=0}^{\psi_{j_D}-1}\frac{(\psi_{j_E})_{r_D}}{r_D!} \sum_{l=1}^{\psi_{j_E}+r_D} \frac{\Gamma(l-\psi_{j_E}-r_D, \mathcal{C}_{j_E}+\mathcal{C}_{j_D})}{\Omega^{\psi_{j_E}-l}_{j_E}\Omega_{j_D}^{r_D-l}(\Omega^{\psi_{j_E}}+\Omega^{\psi_{j_D}})^l}\bigg].
\end{align}

\label{eqn_13}
\begin{align}
I_3&=\sum_{j_E=0}^{M_E}\Lambda_{j_E} e^{\mathcal{C}_{j_E}}\sum_{k=1}^{\psi_{j_E}}\frac{\Gamma(k-\psi_{j_E},\mathcal{C}_{j_E})}{\Omega^{\psi_{j_E}-k}_{j_E}}
\end{align}
where $\Gamma(a,b)=\int_b^\infty x^{a-1} e^{-x} dx$ is the upper incomplete Gamma function [18, eq. (3.351.2)].

\label{App_B}
\begin{IEEEproof}
See Appendix B.
\end{IEEEproof}

\par It can be observed that the ASC has not been presented in [16]. Consequently, to the authors' best knowledge, the expressions in (8)-(13) are novel. 
\section{Secure Outage Probability}
The SOP can be computed by [9, eq. (4)]
\label{eqn_14}
\begin{equation}
\text{SOP}=\int_0^\infty F_D(\theta \gamma_E+\theta-1)f_E(\gamma_E)d\gamma_E
\end{equation}
where $\theta = \mathrm{exp}(R_s) \geq 1$ with $R_s \geq 0$ denotes the target secrecy threshold. 
\par When $\mu$ and $m$ $\in$ $\mathbb{N}^+$, the SOP is expressed as  
\label{eqn_15}
\begin{align}
\text{SOP}&=\frac{\Theta_D \Theta_E}{\mu_D}\frac{\Gamma(\mu_E)}{\Gamma(\mu_E-m_E)}\sum_{i=0}^\infty \sum_{j=0}^\infty \frac{(\theta-1)^{\mu_D+i+j}(-\mathcal{A}_D)^i \mathcal{C}^j_D}{\Gamma(-\mu_D-i-j)(\mathcal{A}_E-\mathcal{B}_E)^{\mu_E}} \nonumber\\
&\times G^{0,1:1,1:1,1}_{1,0:1,1:1,2} 
\Bigg( \begin{matrix}
  1-\mu_E\\
  -\\
\end{matrix} \Vast\vert
\begin{matrix}
  \mu_D+i+j+1\\
  0\\
\end{matrix} \Vast\vert
\begin{matrix}
  1-\mu_E+m_E\\
  0,1-\mu_E\\
\end{matrix} \Vast\vert\frac{\frac{\theta}{\theta-1}}{\mathcal{A}_E-\mathcal{B}_E},\frac{\mathcal{B}_E}{\mathcal{A}_E-\mathcal{B}_E}
\Bigg).
\end{align}
\par For integer values of the fading parameters, the SOP is given in simple exact mathematically tractable closed-from expression as
\label{eqn_16}
\begin{align}
\text{SOP}&=1-\sum_{j_D=0}^{M_D} \Lambda_{j_D} e^{-\frac{\theta-1}{\Omega_{j_D}}} \sum_{r_D=0}^{\psi_{j_D}-1}\frac{1}{\Omega_{j_D}^{r_D} r_D!} \sum_{j_E=0}^{M_E} \frac{\Lambda_{j_E}}{\Omega_{j_E}^{\psi_{j_E}}(\psi_{j_E}-1)!} \nonumber\\
&\times \sum_{k=0}^{r_D}{{r_D}\choose{k}} \frac{\theta^k}{(\theta-1)^{k-r_D}(\frac{\theta}{\Omega_{j_D}}+\frac{1}{\Omega_{j_E}})^{k+\psi_{j_E}}}\Gamma(k+\psi_{j_E})
\end{align}
where ${{b}\choose{a}}\triangleq \frac{b!}{(b-a)!}$ is the binomial coefficient.
 
\label{App_C}
\begin{IEEEproof}
See Appendix C.
\end{IEEEproof}

\par One can see that the SOP in (15) and (16) are new because they have not been derived in the previous works such as [16]. 

\section{Lower Bound of SoP}
According to [6], the SOP$^L$ can be obtained from (14) by inserting $\gamma_E \rightarrow \infty$. Hence, the SOP$^L$ can be calculated by
\label{eqn_17}
\begin{align}
\text{SOP}^L &=\int_0^\infty F_D(\theta \gamma_E) f_E(\gamma_E) d\gamma_E \nonumber\\
& \leq \text{SOP}
\end{align}
\par The $\text{SOP}^L$ for arbitrary numbers of $\mu$ and $m$, the SOP$^L$ can be expressed in terms of a single infinite series as follows
\label{eqn_18}
\begin{align}
\text{SOP}^L&=\Theta_D \Theta_E \theta^{\mu_D}\frac{\Gamma(\mu_D)\Gamma(\mu_E)}{\Gamma(\mu_E-m_E)\Gamma(\mu_D-m_D) \Gamma(m_D)} \sum_{j=0}^\infty \frac{(\theta\mathcal{C}_D)^{j}}{j!}(\mathcal{A}_E-\mathcal{B}_E)^{-(\mu_E+\mu_D+j)}\nonumber\\
&\times G^{0,1:1,2:1,1}_{1,0:2,2:1,2} 
\Bigg( \begin{matrix}
  1-\mu_E-\mu_D-j\\
  -\\
\end{matrix} \Vast\vert
\begin{matrix}
  1-\mu_D+m_D,1-m_D\\
  0,-\mu_D-j\\
\end{matrix} \Vast\vert
\begin{matrix}
  1-\mu_E+m_E\\
  0,1-\mu_E\\
\end{matrix} \Vast\vert\frac{\theta \mathcal{A}_D}{\mathcal{A}_E-\mathcal{B}_E},\frac{\mathcal{B}_E}{\mathcal{A}_E-\mathcal{B}_E}
\Bigg).
\end{align}
\par When the values of both $\mu$ and $m$ are integer, the $\text{SOP}^L$ can be obtained in simple exact closed-from expression as   
\label{eqn_19}
\begin{align}
\text{SOP}^L=1-\sum_{j_E=0}^{M_E} \frac{\Lambda_{j_E}} {\Omega_{j_E}^{\psi_{j_E}}(\psi_{j_E}-1)!} \sum_{j_D=0}^{M_D} \Lambda_{j_D} \sum_{r_D=0}^{\psi_{j_D}-1} \frac{\theta^{r_D}}{\Omega_{j_D}^{\psi_{j_D}} r_D!}\frac{\Gamma(\psi_{j_E}+r_D)}{\big(\frac{\theta}{\Omega_{j_D}}+\frac{1}{\Omega_{j_E}}\big)^{\psi_{j_E}+r_D}}
\end{align}
\label{App_D}
\begin{IEEEproof}
See Appendix D.
\end{IEEEproof}
\section{Probability of Strictly Positive Secrecy Capacity}
The SPSC can be obtained by [9, eq. (12)]
\label{eqn_20}
\begin{equation}
\text{SPSC}=1-\text{SOP} \quad \text{for} \quad \theta=1 
\end{equation}
\par Accordingly, the SPSC for arbitrary and integer values of fading parameters can be deduced from (15) and (16), respectively, after using $\theta = 1$ and plugging the results in (20).

\section{Physical Layer Security over Special Cases of $\kappa-\mu$ Shadowed Fading Model}
The secrecy performance over several main/wiretap fading scenarios can be deduced from the results of the $\kappa-\mu$ shadowed fading by setting the fading parameters for a certain value. For example, when $\kappa \rightarrow 0$, $\mu = 1$, and $m \rightarrow \infty$, $\kappa \rightarrow 0$, $\mu$ = m, $m \rightarrow \infty$, the wireless channels undergo Rayleigh and Nakagami-m fading conditions, respectively. The Rician and Rician shadowed fading models can be obtained by plugging $\kappa = K$, $\mu = 1$, and $m \rightarrow \infty$, $\kappa = K$, $\mu = 1$, $m = m$, respectively. The performance metrics of [10], i.e., the SPSC and the ASC over $\kappa-\mu$ fading can be yielded by substituting $\kappa = K$, $\mu = \mu$, $m \rightarrow \infty$ in the derived expressions.     
\label{fig_1}
\begin{figure}[h!]
\centering
  \includegraphics[width=3.8 in, height=2.74 in]{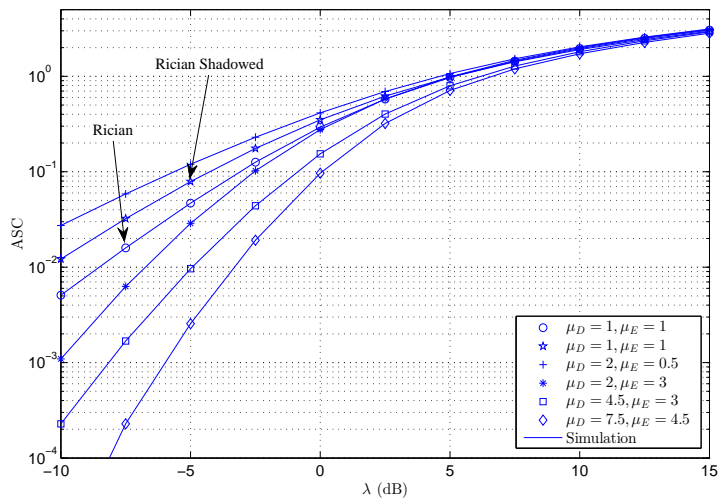} 
\caption{ASC versus $\lambda$ for different values of $\mu_D$, $\mu_E$, $\kappa_D = \kappa_E = 3$, and $m_D = m_E = 2$.}
\end{figure}
\label{fig_2}
\begin{figure}[h!]
\centering
  \includegraphics[width=3.8 in, height=2.74 in]{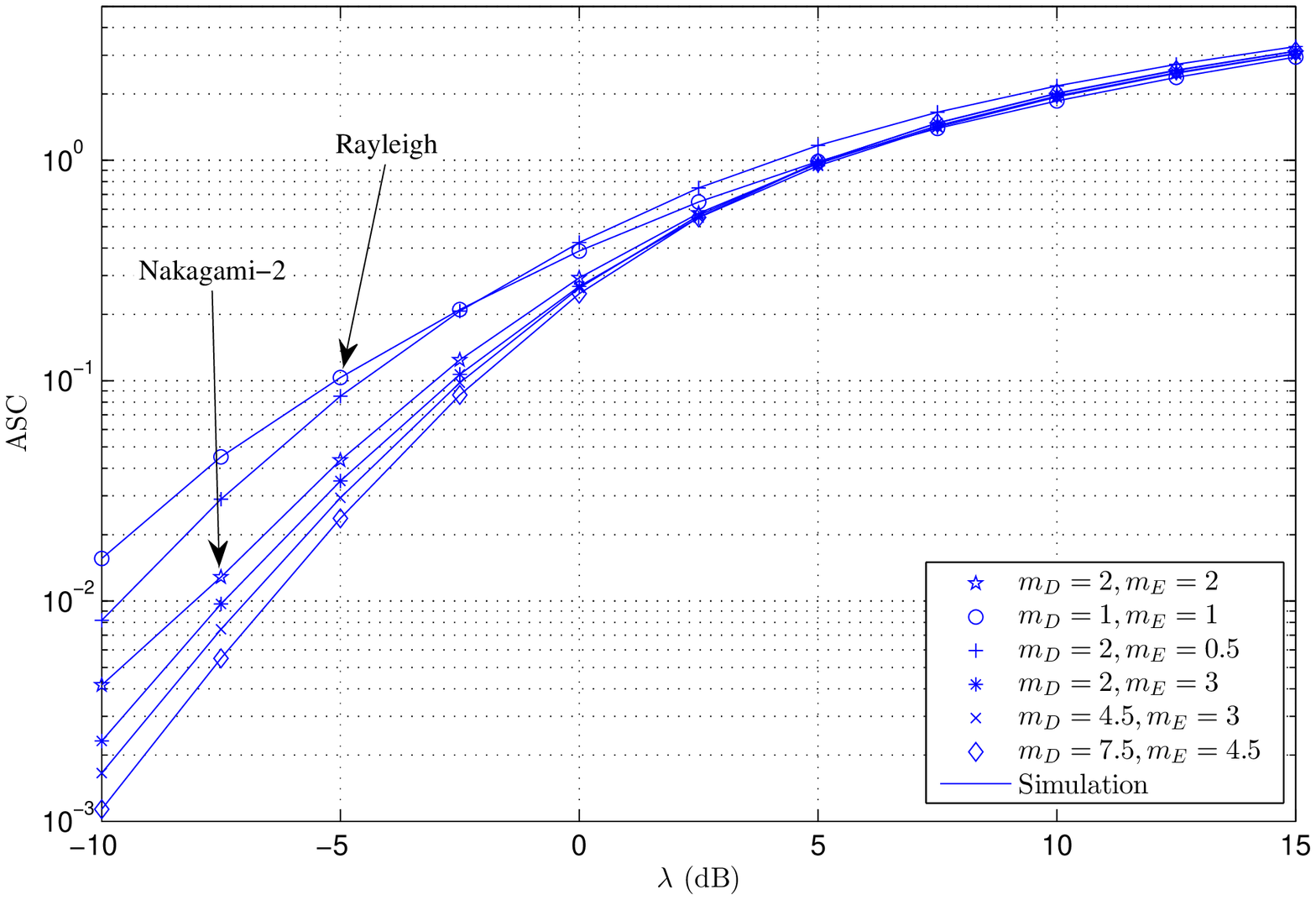} 
\caption{ASC versus $\lambda$ for different values of $m_D$, $m_E$, $\kappa_D = \kappa_E = 3$, and $\mu_D = \mu_E = 2$.}
\end{figure}
\label{fig_3}
\begin{figure}[h!]
\centering
  \includegraphics[width=3.8 in, height=2.74 in]{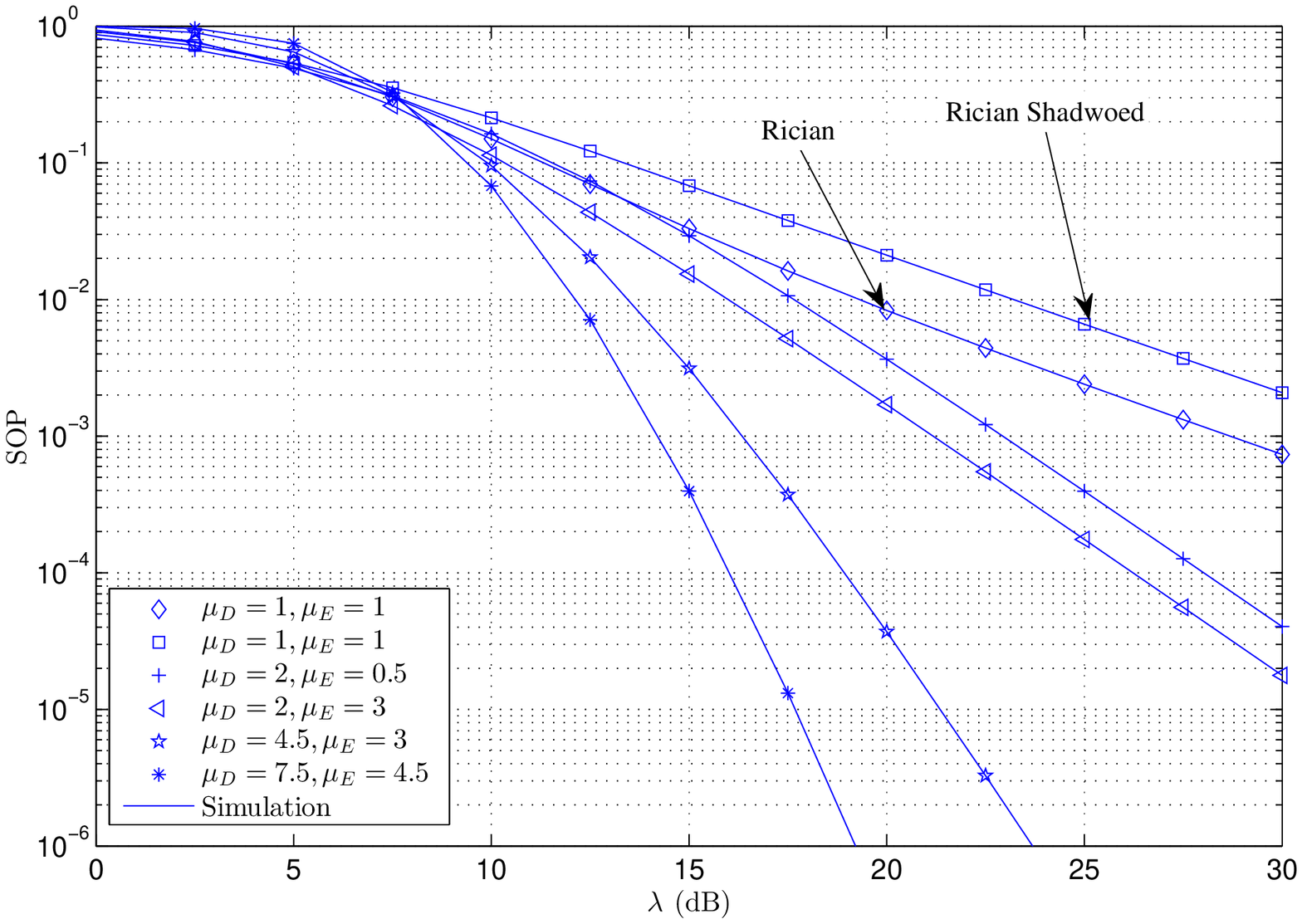} 
\caption{SOP versus $\lambda$ for different values of $\mu_D$, $\mu_E$, $\kappa_D = \kappa_E = 3$, $m_D = m_E = 2$, and $R_s = 1$.}
\end{figure}
\label{fig_4}
\begin{figure}[h!]
\centering
  \includegraphics[width=3.8 in, height=2.74 in]{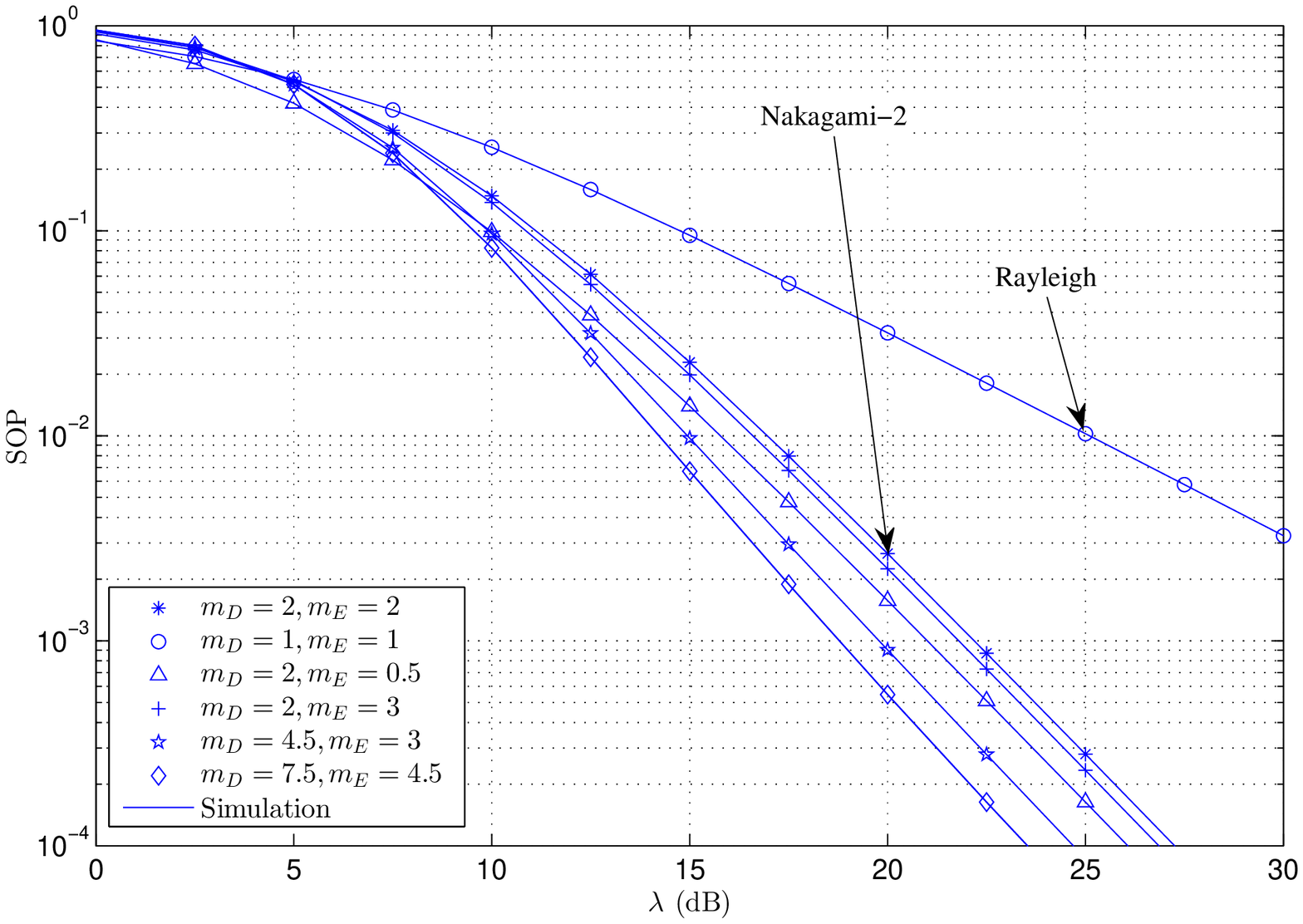} 
 \caption{SOP versus $\lambda$ for different values of $m_D$, $m_E$, $\kappa_D = \kappa_E = 3$, $\mu_D = \mu_E = 2$, and $R_s = 1$.}
 \end{figure} 
 
  \label{fig_5}
  \begin{figure}[h!]
\centering
  \includegraphics[width=3.8 in, height=2.74 in]{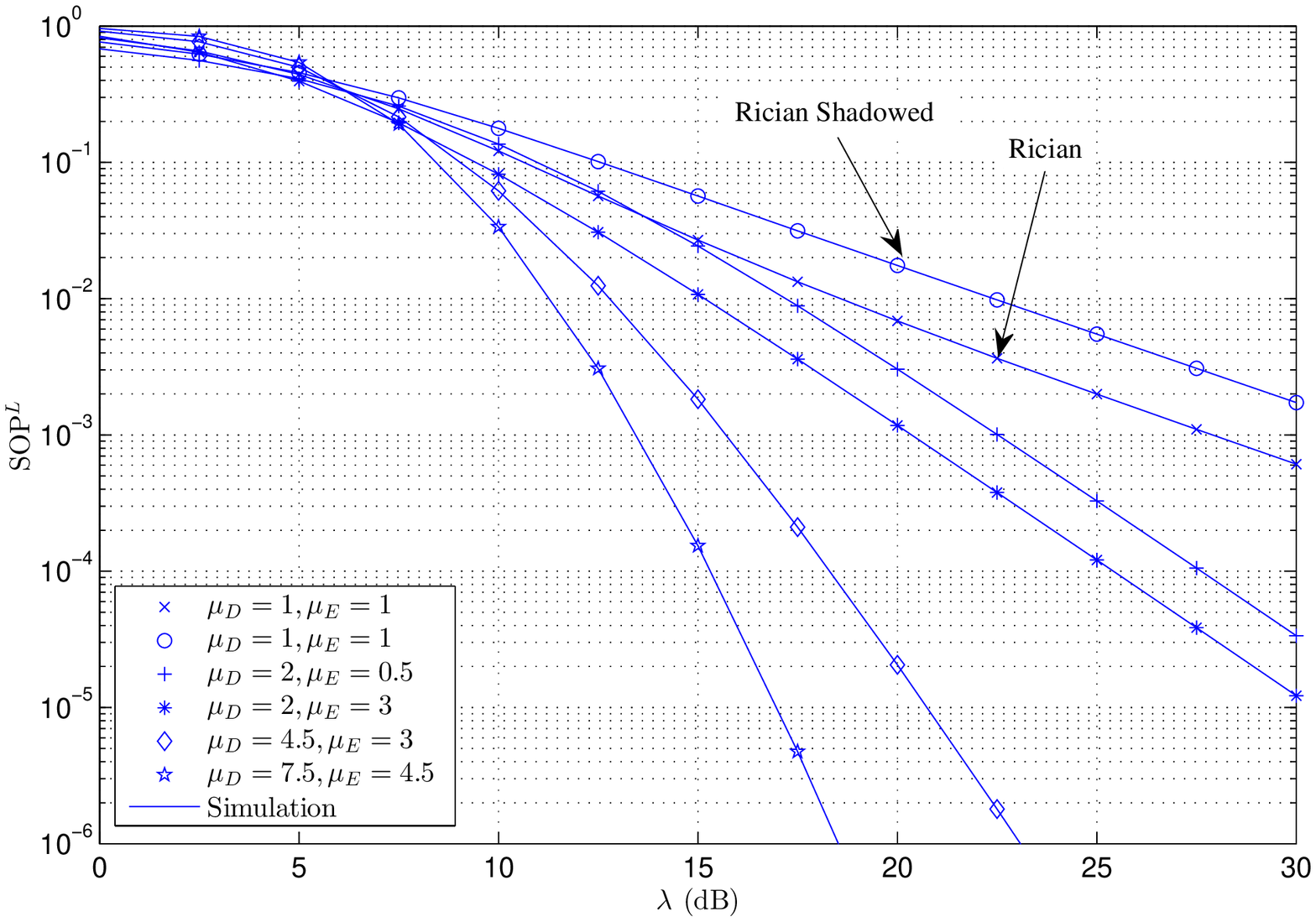} 
\caption{SOP$^L$ versus $\lambda$ for different values of $\mu_D$, $\mu_E$, $\kappa_D = \kappa_E = 3$, $m_D = m_E = 2$, and $R_s = 1$.}
\end{figure}

\label{fig_6}
  \begin{figure}[h!]
\centering
  \includegraphics[width=3.8 in, height=2.74 in]{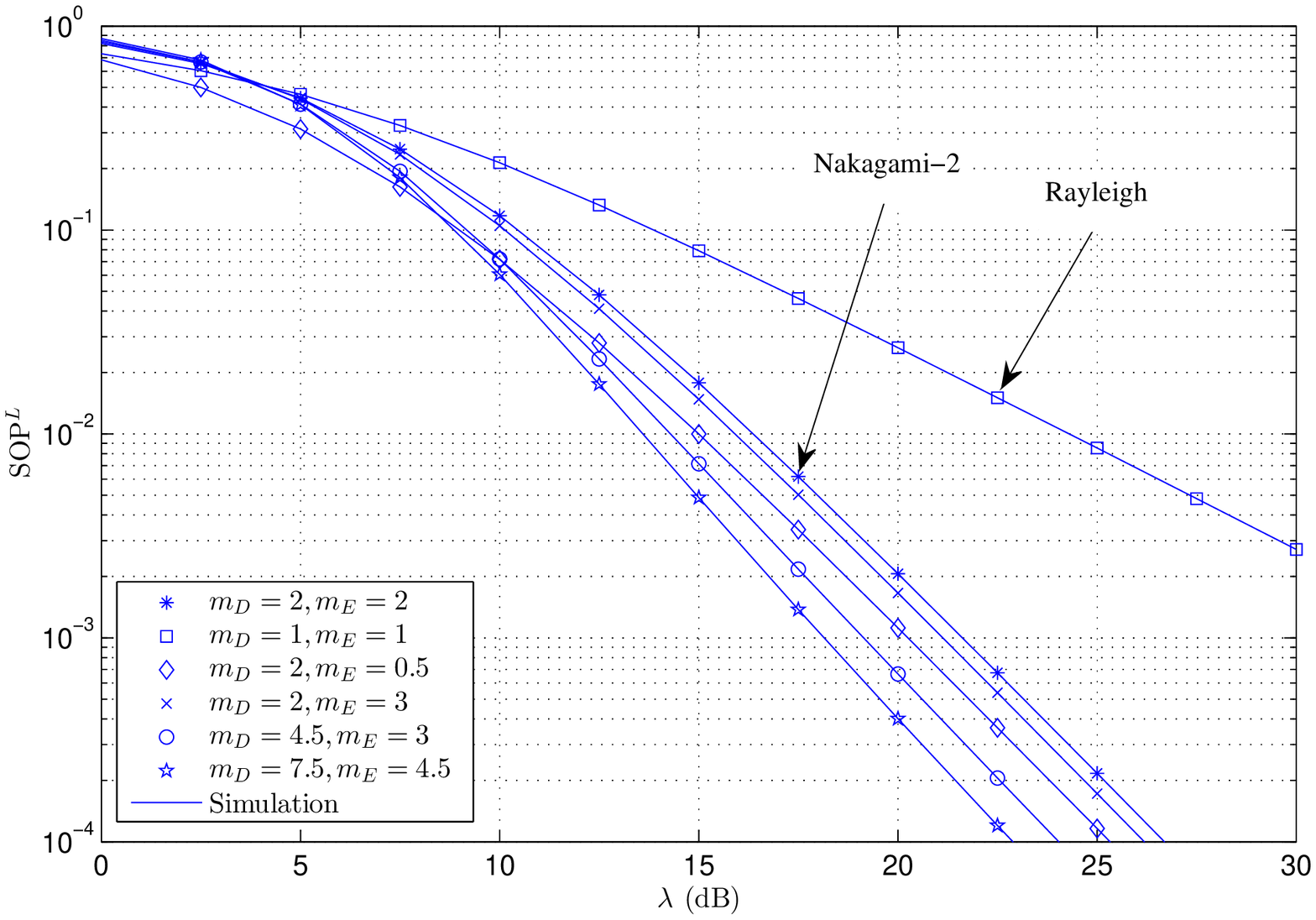}   
 \caption{SOP$^L$ versus $\lambda$ for different values of $m_D$, $m_E$, $\kappa_D = \kappa_E = 3$, $\mu_D = \mu_E = 2$, and $R_s = 1$}
 \end{figure}
  
\label{fig_7}
    \begin{figure}[h!]
\centering
\includegraphics[width=3.8 in, height=2.74 in]{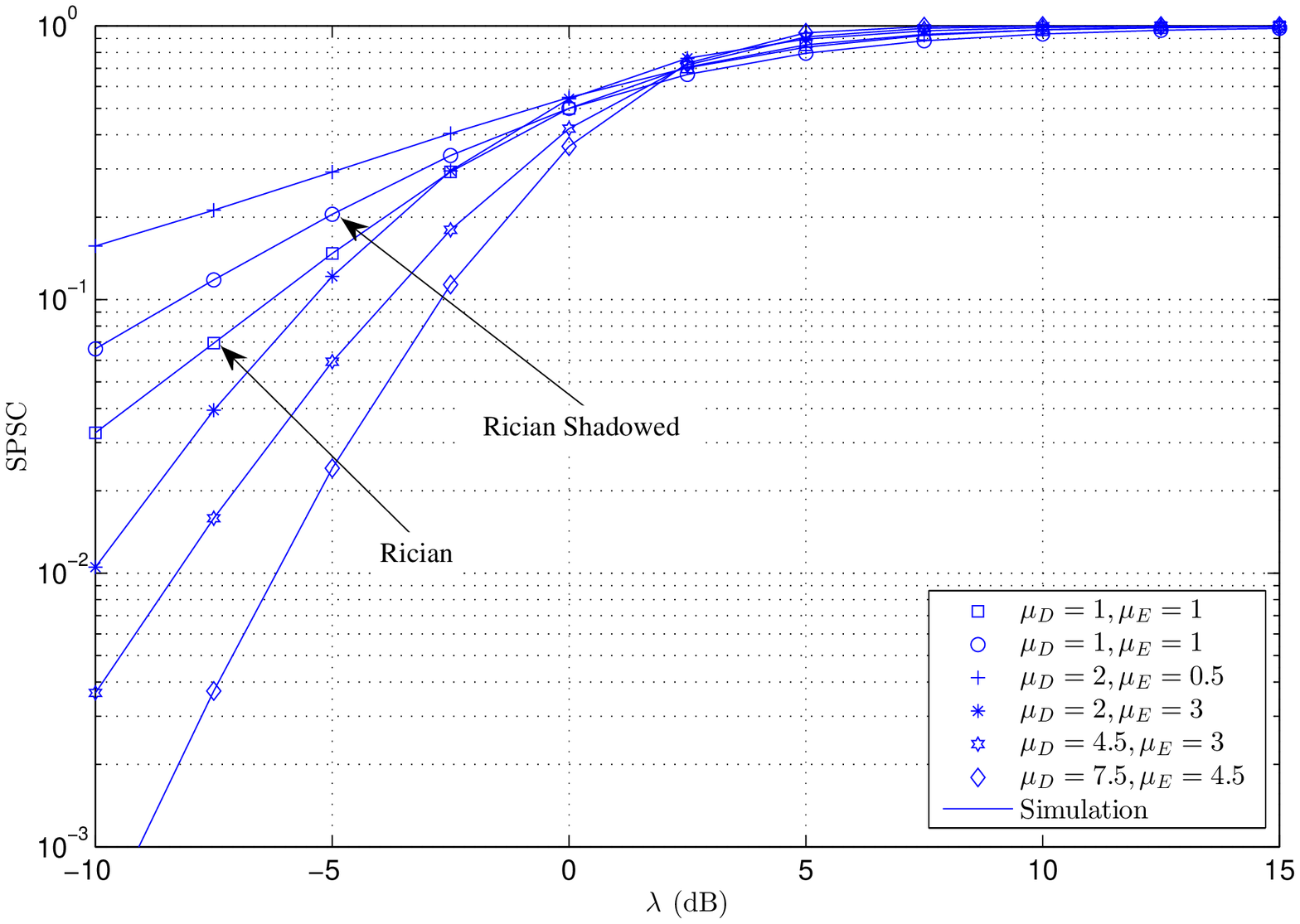}
  \caption{SPSC versus $\lambda$ for different values of $\mu_D$, $\mu_E$, $\kappa_D = \kappa_E = 3$, and $m_D = m_E = 2$.}
 \end{figure}  
  
  \label{fig_8}
    \begin{figure}[h!]
\centering
  \includegraphics[width=3.8 in, height=2.74 in]{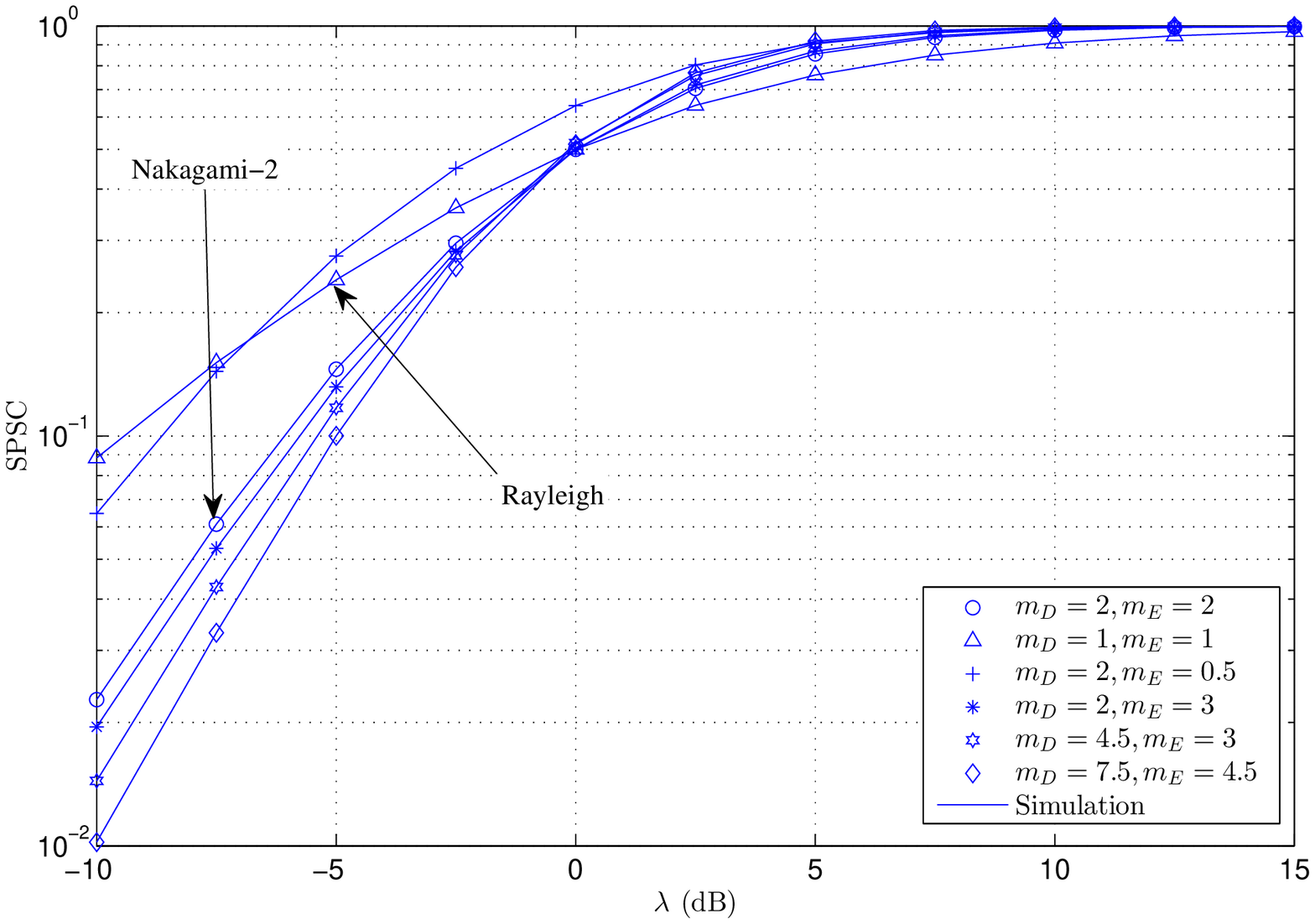} 
 \caption{SPSC versus $\lambda$ for different values of $m_D$, $m_E$, $\kappa_D = \kappa_E = 3$, and $\mu_D = \mu_E = 2$.} 
\end{figure}
\section{Numerical and Simulation Results}
In this section, the numerical results of this work are verified via Monte Carlo simulations with $10^7$ iterations. The parameters of main and wiretap channels are assumed to be independent and non-identically distributed random variables. In all figures, the markers represents the numerical results, whereas the solid lines explain the simulations. Furthermore, all the secrecy performance metrics are plotted versus the ratio $\lambda = \bar{\gamma}_D/\bar{\gamma}_E$ and various values of $\mu$ and $m$. 
\par Figs. 1 and 2 illustrate the ASC for different values of $\mu_i$ and $m_i$ ($i \in \{D, E\}$), respectively, with $\kappa_D = \kappa_E = 3$. From both figures, one can see that the value of the ASC reduces when $\mu_i$ or/and $m_i$ ($i \in \{D, E\}$) increase. This is because the high value of $\mu$ and $m$ refers to the large number of multipath clusters and less shadowing impact at the receiver (Bob or Eve). For instance, in Fig. 1, at $\mu_D = 2$ and $\lambda = -5$ dB (fixed), the ASC for $\mu_E = 0.5$ is nearly $76\%$ high than $\mu_E = 3$. In the same context, when $\mu_E = 3$ and $\mu_D$ changes from $2$ to $4.5$, the ASC is decreased by roughly $66.5\%$. On the other side, in Fig. 2, when $m_D = 2$ at $\lambda = -5$ dB (fixed), the values of the ASC for $m_E = 0.5$ and $m_E = 3$ are approximately $0.085$ and $0.035$, respectively. In addition, the same figure shows the ASC for $m_D = 4.5$ is less than $m_D = 2$ by roughly $16\%$ at constant $\lambda = -5$ dB and $m_E = 3$.  
\par Figs. 3 and 4 demonstrate the SOP for for different values of $\mu_i$ and $m_i$ ($i \in \{D, E\}$), respectively, with $\kappa_D = \kappa_E = 3$, and $R_s = 1$. It can be noticed from these figures that the SOP decreases with the rising of $\mu_i$ or/and $m_i$ ($i \in \{D, E\}$) and for the same reasons that have been mentioned previously. For example, in Fig. 3, at $\lambda = 15$ dB (fixed), the value of the SOP for the ($\mu_D$, $\mu_E$) = ($7.5$, $4.5$) is less by nearly $87\%$ and $98\%$ than ($\mu_D$, $\mu_E$) = ($4.5$, $3$) and ($\mu_D$, $\mu_E$) = ($2$, $0.5$), respectively.  
The provided results in Figs. 3 and 4 are, respectively, affirmed by Figs. 5 and 6 that explain the SOP$^L$ for the same fading and simulation parameters. This confirmation comes from all the results in the latter figures are less than or equal to their corresponding scenarios in Figs. 3 and 4.    
\par Figs. 7 and 8 explain the SPSC for different values of $\mu_i$ and $m_i$ ($i \in \{D, E\}$), respectively, with $\kappa_D = \kappa_E = 3$. As expected, the SPSC decreases with the increasing in the fading parameters ($\mu$ and $m$) of the main or/and the wiretap channels.
\par From all figures, it is clear that the performance improves when the ratio $\lambda$ increases. This refers to the high $\bar{\gamma}_D$ in comparison with the $\bar{\gamma}_E$ which would lead to make the Alice-Bob channel better than the Alice-Eve channel. Moreover, some the performance metrics of physical layer security over special cases of $\kappa-\mu$ shadowed, namely, Rayleigh, Nakagami-$m$, Rician, and Rician shadowed fading channels, have been also investigated. More importantly, the numerical results and Monte Carlo simulations are in perfect match for any provided scenario.

\section{Conclusions}
\par This paper was dedicated to study the secrecy behaviour of the physical layer over $\kappa-\mu$ shadowed fading channels. Different performance metrics, such as the ASC, the SOP, the SOP$^L$, and the SPSC, were derived by assuming two scenarios for the values of the fading parameters, namely,  arbitrary and integer numbers. In the first scenario, the derived results were expressed in terms of EGBMGF and a single infinite series. On the other side, the second scenarios provided simple exact mathematically tractable closed-form expressions. From the given results, a reduction in the values of the ASC, the SOP, the SOP$^L$, and the SPSC can be observed when the values of $\mu$ or/and $m$ increase. Furthermore, the secrecy performances of some special cases of $\kappa-\mu$ shadowed fading channels were also investigated. Accordingly, the results of this work can be employed to study the behaviour of the physical layer over a variety of fading channels with simple exact closed-form expressions and integer fading parameters.         

\section*{Appendix A}
\textbf{\textit{Proof of} (8)}: Substituting (1) and (2) in (5), we have
\label{eqn_21}
\begin{align}
I_1=\frac{\Theta_D \Theta_E}{\mu_E}\int_0^\infty \text{ln}(1+\gamma_D) \gamma^{\mu_D+\mu_E-1}_D e^{-\mathcal{A}_D \gamma_D} {_1F_1(m_D;\mu_D;\mathcal{B}_D \gamma_D)} \nonumber\\
\times\Phi_2\big(\mu_E-m_E,m_E;\mu_E-1;-\mathcal{A}_E \gamma_D,\mathcal{C}_E \gamma_D \big)d\gamma_D.
\end{align}   
\par With the help of [21, eq. (1.3.7)], [18, eq. (9.261.2)], [21, eq. (1.1.20)], and [18, eq. (9.14.1)], we have
\label{eqn_22}
\begin{align}
I_1=&\Theta_D \Theta_E\frac{\Gamma(\mu_D)\Gamma(\mu_E)}{\Gamma(\mu_D-m_D)\Gamma(\mu_E-m_E)\Gamma(m_E)}\sum_{j=0}^\infty\frac{\mathcal{C}^j_E}{j!}\nonumber\\
&\times \int_0^\infty \text{ln}(1+\gamma_D) \gamma^{\mu_D+\mu_E+j-1}_D e^{-(\mathcal{A}_D-\mathcal{B}_D) \gamma_D}{_1F_1(\mu_D-m_D;\mu_D;-\mathcal{B}_D \gamma_D)} \nonumber\\
&\times {_2F_1(\mu_E-m_E,\mu_E;\mu_E+j+1;-\mathcal{B}_D \gamma_D)} 
 d\gamma_D.
\end{align}
\par Using the identities [22, eq. (11)], [18, eq. (9.34.8)], and [18, eq. (9.34.7)] to express $e^{-x}$ and $\text{ln}(1+x)$, ${_1F_1(x;y;-z)}$, and ${_2F_1(x_1,x_2;y;-z)}$ in terms of Meijer $G$-function. Thus, (22) can be rewritten as
\label{eqn_23}
\begin{align}
I_1=&\Theta_D \Theta_E\frac{\Gamma(\mu_D)\Gamma(\mu_E)}{\Gamma(\mu_D-m_D)\Gamma(\mu_E-m_E)\Gamma(m_E)}\sum_{j=0}^\infty\frac{\mathcal{C}^j_E}{j!}\nonumber\\
&\times  
\int_0^\infty \gamma^{\mu_D+\mu_E+j-1}_D G^{1,2}_{2,2} \Bigg(\begin{matrix}
  1,1\\
  1,0\\
\end{matrix} \Vast\vert
 \gamma_D\Bigg) 
 G^{1,0}_{0,1} \Bigg( \begin{matrix}
  -\\
  0\\
\end{matrix} \Vast\vert
 (\mathcal{A}_D-\mathcal{B}_D)\gamma_D\Bigg)\nonumber\\
&\times  
 G^{1,1}_{1,2} \Bigg( \begin{matrix}
  1-\mu_D+m_D\\
  0,1-\mu_D\\
\end{matrix} \Vast\vert
 \mathcal{B}_D\gamma_D\Bigg)
G^{1,2}_{2,2} \Bigg( \begin{matrix}
  1-\mu_E-m_E,1-m_E\\
  0,-\mu_E-j\\
\end{matrix} \Vast\vert
 \mathcal{A}_E\gamma_D\Bigg)
 d\gamma_D.
\end{align}
\par The integral in (23) can be computed by [23, eq. (9)] and that completes the proof of (8).
\par \textbf{\textit{Proof of} (9)}: It can be observed that (9) can be calculated by (8) after replacing the symbols $D$ and $E$ with $E$ and $D$, respectively.
\par \textbf{\textit{Proof of} (10)}: Plugging (1) in (7), we have 
\label{eqn_24}
\begin{align}
I_3=\Theta_E \int_0^\infty \text{ln}(1+\gamma_E) \gamma^{\mu_E-1}_E e^{-\mathcal{A}_E \gamma_E} {_1F_1(m_E;\mu_E;\mathcal{B}_E \gamma_E)}d\gamma_E
\end{align}
\par Similar to (9), the identities [22, eq. (11)] and [18, eq. (9.34.8)] are utilised to obtain
\label{eqn_25}
\begin{align}
I_3&=\Theta_E \frac{ \Gamma(\mu_E)}{\Gamma(\mu_E-m_E)} \int_0^\infty \gamma^{\mu_E-1}_E G^{1,2}_{2,2} \Bigg(\begin{matrix}
  1,1\\
  1,0\\
\end{matrix} \Vast\vert
 \gamma_E
\Bigg) \nonumber\\ 
& \times G^{1,0}_{0,1} \Bigg( \begin{matrix}
  -\\
  0\\
\end{matrix} \Vast\vert
 (\mathcal{A}_E-\mathcal{B}_E)\gamma_E
\Bigg)
G^{1,1}_{1,2} \Bigg( \begin{matrix}
  1-\mu_E+m_E\\
  0,1-\mu_E\\
\end{matrix} \Vast\vert
 \mathcal{B}_E\gamma_E
\Bigg)d\gamma_E.
\end{align}
\par With the aid of [23, eq. (9)], (25) can be evaluated in exact closed-form expression as in (10).
\section*{Appendix B}
\par \textbf{\textit{Proof of} (11)}: After inserting (3) and (4) in (5), this yields
\label{eqn_26}
\begin{align}
I_1&=\sum_{j_D=0}^{M_D}\frac{\Lambda_{j_D}}{\Omega^{\psi_{j_D}}_{j_D} (\psi_{j_D}-1)!}\bigg[\int_0^\infty \text{ln}(1+\gamma_D) \gamma^{\psi_{j_D}-1}_D e^{-\frac{\gamma_D}{\Omega_{j_D}}} d\gamma_D\nonumber\\
&-\sum_{j_E=0}^{M_E}\Lambda_{j_E}\sum_{r_E=0}^{\psi_{j_E}-1}\frac{1}{\Omega_{j_E}^{r_E} r_E!}\int_0^\infty \text{ln}(1+\gamma_D) \gamma^{\psi_{j_D}+r_E-1}_D e^{-\big(\frac{1}{\Omega_{j_D}}+\frac{1}{\Omega_{j_E}}\big)\gamma_D} d\gamma_D\bigg].
\end{align}
\par Employing [24, eq. (47)] to calculate the integrals of (26) in simple exact closed-form expressions as given in (11).
\par \textbf{\textit{Proof of} (12)}: Using $D$ and $E$ instead of $E$ and $D$, respectively, in (11), the result is $I_2$ that is given in (12).
\par \textbf{\textit{Proof of} (13)}: Plugging (3) in (7), this yields
\label{eqn_27} 
\begin{align}
I_3&=\sum_{j_E=0}^{M_E}\frac{\Lambda_{j_E}}{\Omega^{\psi_{j_E}}_{j_E} (\psi_{j_E}-1)!}\int_0^\infty\text{ln}(1+\gamma_E)\gamma^{\psi_{j_E}-1}_E e^{-\frac{\gamma_E}{\Omega_{j_E}}}d\gamma_E.
\end{align}
\par Likewise, [24, eq. (47)] is utilised to express (27) in exact closed-form as in (13).
\section*{Appendix C}
\par \textbf{\textit{Proof of} (15)}: Inserting (1) and (2) in (14) and using [18, eq. (9.261.2)] and [21, eq. (1.3.7)], the result is
\label{eqn_28}
\begin{align}
\text{SOP}&=\frac{\Theta_D \Theta_E}{\mu_D}\sum_{i=0}^\infty \sum_{j=0}^\infty (-\mathcal{A}_D)^i \mathcal{C}^j_D \nonumber\\
&\times\int_0^\infty (\theta \gamma_E+\theta-1)^{\mu_D+i+j} \gamma^{\mu_E-1}_E e^{-(\mathcal{A}_E-\mathcal{B}_E)\gamma_E} {_1F_1}(\mu_E-m_E;\mu_E;-\mathcal{B}_E \gamma_E) d\gamma_E
\end{align}
\par After doing some mathematical manipulations and recalling the identities [22, eq. (10)], [18, eq. (9.34.8)], and [22, eq. (17)], to write $(1+x)^a$, $e^{-x}$, and $_1F_1(x;y;-z)$, respectively, using Meijer $G$-function, (28) becomes 
\label{eqn_29}
\begin{align}
\text{SOP}&=\frac{\Theta_D \Theta_E}{\mu_D}\frac{\Gamma(\mu_E)}{\Gamma(\mu_E-m_E)}\sum_{i=0}^\infty \sum_{j=0}^\infty \frac{(\theta-1)^{\mu_D+i+j}(-\mathcal{A}_D)^i \mathcal{C}^j_D}{\Gamma(-\mu_D-i-j)} \nonumber\\
&\times\int_0^\infty \gamma^{\mu_E-1}_E G^{1,0}_{0,1} \Bigg(\begin{matrix}
  -\\
  0\\
\end{matrix} \Vast\vert
 (\mathcal{A}_E-\mathcal{B}_E)\gamma_E
\Bigg) \nonumber\\
 &\times G^{1,1}_{1,1} \Bigg( \begin{matrix}
  \mu_D+i+j+1\\
  0\\
\end{matrix} \Vast\vert
 \frac{\theta}{\theta-1}\gamma_E
\Bigg)
G^{1,1}_{1,2} \Bigg( \begin{matrix}
  1-\mu_E+m_E\\
  0,1-\mu_E\\
\end{matrix} \Vast\vert
 \mathcal{B}_E\gamma_E
\Bigg)d\gamma_E.
\end{align}
\par With the aid of [23, eq. (9)], the expression of the SOP in (29) can be deduced as in (15) which completes the proof. 
\par \textbf{\textit{Proof of} (16)}: Substituting (3) and (4) in (14) and utilising $\int_0^\infty f_\gamma(\gamma) d\gamma \triangleq 1$, we have
\label{eqn_30}
\begin{align}
\text{SOP}&=1-\sum_{j_D=0}^{M_D} \Lambda_{j_D} e^{-\frac{\theta-1}{\Omega_{j_D}}} \sum_{r_D=0}^{\psi_{j_D}-1}\frac{1}{\Omega_{j_D}^{r_D} r_D!} \sum_{j_E}^{M_E} \frac{\Lambda_{j_E}}{\Omega_{j_E}^{\psi_{j_E}}(\psi_{j_E}-1)!} \nonumber\\
&\times \int_0^\infty \gamma_E^{\psi_{j_E}-1} (\theta \gamma_E+\theta-1)^{r_D} e^{-\big(\frac{\theta}{\Omega_{j_D}}+\frac{1}{\Omega_{j_E}}\big)\gamma_E}d\gamma_E
\end{align}
\par Doing some mathematical manipulations with the aid $(1+a)^b=\sum_{k=0}^{b}{{b}\choose{k}}a^k$ [18, eq. (1.111)] and employing [18, eq. (3.381.4)], the integral of (30) can be expressed in simple exact closed-form as in (16).
\section*{Appendix D}
\par \textbf{\textit{Proof of} (18)}: Inserting (1) and (2) in (17) and following the same procedure for (28), this yields
\label{eqn_31}
\begin{align}
\text{SOP}^L&=\Theta_D \Theta_E \theta^{\mu_D}\frac{\Gamma(\mu_D)\Gamma(\mu_E)}{\Gamma(\mu_E-m_E)\Gamma(\mu_D-m_D) \Gamma(m_D)} \sum_{j=0}^\infty \frac{(\theta\mathcal{C}_D)^{j}}{j!}\nonumber\\
&\times\int_0^\infty \gamma^{\mu_E+\mu_D+j-1}_E G^{1,0}_{0,1} \Bigg(\begin{matrix}
  -\\
  0\\
\end{matrix} \Vast\vert
 (\mathcal{A}_E-\mathcal{B}_E)\gamma_E
\Bigg)  \nonumber\\
&\times G^{1,2}_{2,2} \Bigg( \begin{matrix}
  1-\mu_D+m_D,1-m_D\\
  0,-\mu_D-j\\
\end{matrix} \Vast\vert
 \theta\mathcal{A}_D\gamma_E
\Bigg)
G^{1,1}_{1,2} \Bigg( \begin{matrix}
  1-\mu_E+m_E\\
  0,1-\mu_E\\
\end{matrix} \Vast\vert
 \mathcal{B}_E\gamma_E
\Bigg)d\gamma_E.
\end{align}
\par Again, [23, eq. (9)] is employed to evaluate the integral of (31) as given in (18) which completes the proof.
\par \textbf{\textit{Proof of} (19)}: Plugging (3) and (4) in (17) and employing $\int_0^\infty f_\gamma(\gamma) d\gamma \triangleq 1$, we have
\label{eqn_33}
\begin{align}
\text{SOP}^L=1-\sum_{j_E=0}^{M_E} \frac{\Lambda_{j_E}} {\Omega_{j_E}^{\psi_{j_E}}(\psi_{j_E}-1)!} \sum_{j_D=0}^{M_D} \Lambda_{j_D} \sum_{r_D=0}^{\psi_{j_D}-1} \frac{\theta^{r_D}}{\Omega_{j_D}^{\psi_{j_D}} r_D!}\int_0^\infty \gamma_E^{\psi_{j_E}+r_D-1}e^{-\big(\frac{\theta}{\Omega_{j_D}}+\frac{1}{\Omega_{j_E}}\big)\gamma_E}d\gamma_E
\end{align}
\par With the aid of [18, eq. (3.381.4)], (32) can be calculated in simple exact closed-form expression as provided in (19).  

\end{document}